\documentclass[%
 aip,
 amsmath,amssymb,floatfix,
 reprint%
]{revtex4-1}

\usepackage{graphicx}
\usepackage{dcolumn}
\usepackage{bm}

\usepackage[utf8]{inputenc}
\usepackage[T1]{fontenc}
\usepackage{mathptmx}
\usepackage{etoolbox}
\usepackage{subfigure}
\usepackage{hyperref}
\usepackage{siunitx}
\usepackage[final]{changes}

\newcommand{\ub}{{\bf u}}

\newcommand{\Xb}{{\bf X}}

\newcommand{\xb}{{\bf x}}

\newcommand{\fb}{{\bf f}}

\newcommand{\partialb}{{\boldsymbol \partial}}

\makeatletter
\def\@email#1#2{%
 \endgroup
 \patchcmd{\titleblock@produce}
  {\frontmatter@RRAPformat}
  {\frontmatter@RRAPformat{\produce@RRAP{*#1\href{mailto:#2}{#2}}}\frontmatter@RRAPformat}
  {}{}
}%
\makeatother
\begin{document}

\preprint{AIP/123-QED}

\title{Universal flapping states of elastic fibers in modulated turbulence}
\author{Stefano Olivieri}
\affiliation{
 Complex Fluids and Flows Unit, Okinawa Institute of Science and Technology Graduate University, 1919-1 Tancha, Onna-son, Okinawa 904-0495, Japan
}

\author{Andrea Mazzino}
\affiliation{
 Department of Civil, Chemical and Environmental Engineering (DICCA), University of Genova, Via Montallegro 1, 16145 Genova, Italy
}
\affiliation{
 INFN, Genova Section, Via Montallegro 1, 16145 Genova, Italy
}

\author{Marco E. Rosti}
\affiliation{
 Complex Fluids and Flows Unit, Okinawa Institute of Science and Technology Graduate University, 1919-1 Tancha, Onna-son, Okinawa 904-0495, Japan
}

\email[Corresponding author:]{marco.rosti@oist.jp}

\date{\today}

\begin{abstract}
We study the fully-coupled dynamics between a fully-developed turbulent flow and an ensemble of immersed flexible fibers. We vary the concentration of the suspension, the mechanical properties and the length of the fibers in a vast parametric range. For all configurations, the fiber dynamics falls in only two possible dynamical states: (i) the fiber manifests its natural response to the flow forcing or (ii) its motion fully synchronizes to the hydrodynamic timescales of the turbulent flow. This scenario holds for both a dilute condition, where the carrier flow is not affected by the fluid-structure interaction, as well as in the case where the flow is substantially altered by the presence of the immersed objects. Such backreaction effect can be macroscopically modelled in terms of the mass fraction of the suspension. Our results can be readily extended to any elastic objects interacting with fluid turbulence.
\end{abstract}

\maketitle

The interaction between elastic objects and fluid flows plays an essential role in a variety of both fundamental and applied problems, ranging from biological  (e.g., mucus-cilia or red blood cells hydrodynamics)~\cite{takeishi2017capture,duroure2019review,loiseau2020active} and environmental processes (e.g., flow in flexible canopies and plant reconfiguration)~\cite{delangre2008effects,abdolahpour2018impact,monti2020genesis,tschisgale2021large}  to engineering applications (e.g., energy harvesting and flow control)\added{~\cite{hobeck2012artificial,orrego2017harvesting,olivieri2017fluttering,boccalero2017power,mackenzie2018,olivieri2019constructive,rosti2020low}}.  Despite the great advancement of research in these fields, the available knowledge on such nontrivial phenomena mostly concerns the situation where the flow is laminar\added{~\cite{bagheri2012spontaneous,gsell2020hydrodynamic,deng2021symmetry,andersson2021}}, whereas the interaction between elastic objects and turbulence has started to be understood only more recently~\cite{brouzet2014flexible,rosti2018flexible,rosti2019flowing,picardo2020dynamics,zuk2021}. In this framework, naturally arising questions that are still open are: Does a universal scenario hold in the interaction between an ensemble of elastic objects and a turbulent flow? And what are the parameters controlling such flow-structure interaction?

Focusing on very dilute suspensions of flexible fibers in a homogeneous turbulent flow, Refs.~\cite{rosti2018flexible,rosti2019flowing} provided novel insight by means of a simple phenomenological model based on the characteristic timescales (i.e., structural, viscous and hydrodynamic) that can be identified in the problem. Using an analogy with a damped harmonic oscillator, it is therefore possible to distinguish two dynamical regimes (i.e., underdamped vs overdamped) and further predict different fiber flapping states~\cite{rosti2019flowing}.  However, the main limitation of such original approach is the implicit \textit{one-way} coupling assumption, i.e. the suspension is considered to be dilute enough so that the backreaction of the dispersed phase to the carrier flow can be safely neglected. On the other hand, the backreaction effect has been recently investigated for fixed objects by Refs.~\cite{olivieri2020dispersed,olivieri2020turbulence} revealing a robust nonlocal energy transfer (from larger to smaller flow scales). This effect turns out to be not negligible in the non-dilute condition, causing the strong modulation of the fluid flow across all scales of motion.  

\begin{figure}[b]
    \centering
    \includegraphics{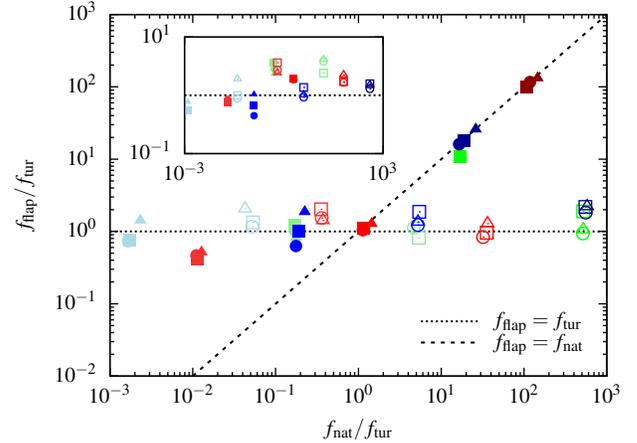}
    \caption{\added{Fiber flapping frequency as a function of the natural frequency, both normalized by the \emph{effective} eddy turnover frequency $f_\mathrm{tur} =  \beta \, \sqrt{S_2} / c$ (see the definition in main text). The two possible flapping states, i.e. $f_\mathrm{flap} = f_\mathrm{tur}$ or $f_\mathrm{flap} = f_\mathrm{nat}$, are indicated by the dotted and dashed line, respectively, and compared with the results from our DNS study. Filled: $\Delta \widetilde{\rho} / (\rho_\mathrm{f} L^2) \approx 0.025$ (i.e., inertial); empty: $\Delta \widetilde{\rho} / (\rho_\mathrm{f} L^2) \ll 1$ (i.e, neutrally-buoyant). Symbols and colors respectively denote different concentration (circle: $N=10^1$; square: $N=10^2$; triangle: $N=10^3$) and length (green: $c/L \approx 0.016$; red: $c/L \approx 0.08$; blue: $c/L \approx 0.32$) while the bending stiffness varies according to brightness (light: $\gamma/\gamma_0 \approx 10^{-4}$; medium: $\gamma/\gamma_0 \approx 1$; dark: $\gamma/\gamma_0 \approx 10^{4}$, with $\gamma_0 = \rho_\mathrm{f} (\nu L)^2$).
    The inset shows the different data scaled by the (single-phase) large-eddy turnover frequency $f_\mathrm{tur}=u_\mathrm{rms}/L$.}}
    \label{fig:flapFreq}
\end{figure}

\begin{figure*}[t]
    \centering
    \includegraphics[width=.95\textwidth]{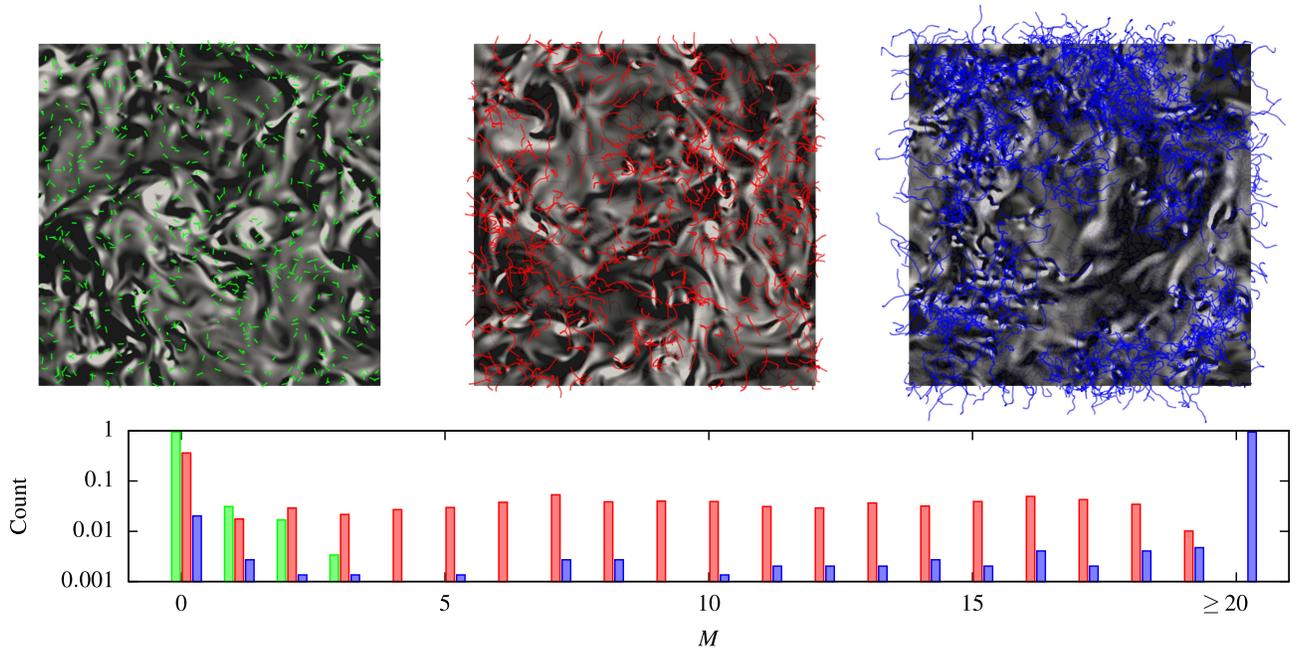}
    \caption{(Top) Sideview snapshots from DNS of $N=10^3$ inertial fibers with $\gamma/\gamma_0 \approx 1$ dispersed in homogeneous isotropic turbulence for three different fiber lengths $c/L \approx (0.016, 0.08, 0.32)$,  showing the normal component of the vorticity field in a cutplane (negative values colored in white, positive values in black) along with the instantaneous fiber positions.  (Bottom) Normalized count of inflection points for the same cases (with $M=0$ indicating a straight fiber and limited to $M=20$).}
    \label{fig:snapshots}
\end{figure*}

This Letter investigates the fully-coupled dynamics between a homogeneous turbulent flow and a suspension of flexible fibers in a non-dilute condition, such that the flow is significantly altered by the presence of the dispersed phase.  \added{As anticipated in Fig.~\ref{fig:flapFreq}}, we find that only two flapping states are possible depending on the ratio between the natural frequency $f_\mathrm{nat}$ and the hydrodynamic frequency at the fiber lengthscale $f_\mathrm{tur}$: fibers are either intermittently manifesting an essentially free response at their natural frequency, i.e. $f_\mathrm{flap} \approx f_\mathrm{nat}$ (where $f_\mathrm{flap}$ is the dominant flapping frequency) or are locked to the turbulent structures of the modified fluid flow, i.e. $f_\mathrm{flap} \approx f_\mathrm{tur}$.  Since it is governed only by the balance between effective timescales, this universal scenario is expected to hold for any problem involving the mutual interaction between fluid turbulence and elastic objects.
\added{
To achieve the accurate data rescaling shown in Fig.~\ref{fig:flapFreq}, two aspects are key (and will be further discussed in the following of this Letter): (i) the hydrodynamic frequency here identified is that of turbulent eddies of the same size of the fiber; (ii) such characteristic timescale has to properly account for the modulation of the flow caused by the backreaction effect.
}

\added{Our results} are obtained by means of direct numerical simulations (DNS) in a cubic domain of size $L$ of homogeneous isotropic turbulence combined with an immersed boundary method (IBM) to solve the fully-coupled flow-structure interaction~\footnote{See Supplemental Material for more information on the physical model, numerical method and performed analysis.}. In particular, we consider an incompressible Newtonian flow governed by the well-known Navier-Stokes equations
\begin{equation}
\begin{aligned}
\partial_t \ub + \ub \cdot \partialb \ub &= - \partialb p / {\rho_\mathrm{f}} + \nu \partial^2 \ub + \fb_\mathrm{tur} +  \fb_\mathrm{fib},\\
\partialb \cdot \ub &= 0,
\label{eq:NS}
\end{aligned}
\end{equation}
where $\ub \left( \xb,t \right)$ and $p\left( \xb,t \right)$ are the fluid velocity and pressure fields, $\rho_\mathrm{f}$ and $\nu$ are the fluid volumetric density and kinematic viscosity, $\fb_\mathrm{tur}$ and $\fb_\mathrm{fib}$ are the forcings used to sustain turbulence and model the presence of the suspended fibers, respectively~\cite{rosti2018flexible,rosti2019flowing,olivieri2020dispersed,olivieri2020turbulence,cavaiola2019assembly}.  Each dispersed fiber is modeled as an inextensible elastic filament,
\begin{equation}
\begin{aligned}
\Delta \widetilde{\rho} \,  \ddot{\Xb}&= \partial_s \left( T \partial_s (\Xb) \right) - \gamma \partial^4_s \Xb - \textbf{F},\\
\partial_s \Xb \cdot \partial_s \Xb &= 1,
\end{aligned}
\end{equation}
where $\textbf{X} \left( s, t \right)$ is the position of a given material point as a function of the curvilinear coordinate $s$ and time $t$, $\Delta \widetilde{\rho} = \widetilde{\rho}_\mathrm{s} - \widetilde{\rho}_\mathrm{f}$ is the linear density difference between the fiber and the fluid, $T$ is the tension enforcing the inextensibility, $\gamma$ is the bending stiffness and $\textbf{F} \left( s, t \right)$ is the fluid-structure interaction force.  From a normal-mode analysis in the case of dispersed fibers (i.e. free-free unsupported configuration) we obtain the natural frequency $f_\mathrm{nat} = \alpha \, \sqrt{\gamma / (\Delta \widetilde{\rho} c^4)}$ (where $\alpha \approx 22.4/\pi$).

The universality of our findings has been proved by performing a vast parametric study considering fibers of different length $c$ (from short fibers comparable to the dissipative lengthscale to long fibers comparable to the integral lengthscale), and mechanical properties, i.e. bending stiffness $\gamma$ (from essentially rigid to very deformable fibers), linear density difference $\Delta \widetilde{\rho}$ (from almost neutrally-buoyant to heavy fibers), along with varying the number of fibers $N$ (from dilute to dense suspensions).  An example of the extremely different configurations that have been analyzed is shown in Fig.~\ref{fig:snapshots} (top panels) for fibers of three different lengths (while keeping the same mechanical properties): (i) short fibers comparable with the Kolmogorov lengthscale (left panel, green fibers), (ii) intermediate fibers lying within the inertial subrange (center, red), (iii) long fibers comparable to the integral lengthscale (right, blue). Fibers deform very differently, with the shortest remaining almost straight and the longest exhibiting several inflections. A quantitative evaluation of the fiber deformation is shown in the bottom panel of Fig.~\ref{fig:snapshots} reporting the number of inflection points $M$ experienced on average by the fibers: the range of activated modes clearly gets broader while increasing the fiber length, with short fibers found almost exclusively in the straight configuration ($M=0$), intermediate ones with a more wider distribution, and a further more complex configuration for the longest fibers. Finally, from the top panels of Fig.~\ref{fig:snapshots} we observe the different tendency to form clusters while varying the fiber length and/or concentration.
 
 \begin{figure}[t]
    \centering
    \includegraphics{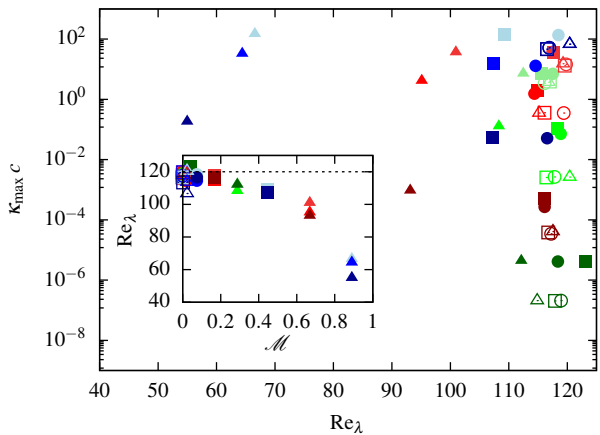}
    \caption{Fiber maximum curvature (made nondimensional with the fiber length $c$) vs flow micro-scale Reynolds number. The legend is the same of Fig.~\ref{fig:flapFreq}. Filled: inertial fibers; empty: neutrally-buoyant fibers. Symbols denote different concentration (circle: low; square: medium; triangle: high). Colors denote different fiber lengths (green: short; red: intermediate; blue: long). The bending stiffness is varied according to the brightness (increasing from light to dark). The inset shows the dependency of the turbulence modulation on the mass fraction, with the horizontal dashed line indicating the micro-scale Reynolds number in the single-phase case.}
    \label{fig:Kmaxc-vs-ReL}
\end{figure}

Not only the fiber deformation varies substantially, but also the resulting fluid flow differs appreciably when changing the suspension properties. This is shown quantitatively in Fig.~\ref{fig:Kmaxc-vs-ReL} relating the fiber deformation, expressed in terms of the maximum fiber curvature $\kappa_\mathrm{max}$, and a bulk property of the flow, the micro-scale Reynolds number $\mathrm{Re}_\lambda$.  The two quantities can be used together to describe the key aspects of the structural and fluid dynamics, respectively.  Looking at the variation of the fluid flow due to the backreaction (along the horizontal axis), a clear distinction can be made between the inertial and the neutrally-buoyant fibers. For the inertial fibers we have that $\mathrm{Re}_\lambda$ changes dramatically while varying the fiber concentration, reducing by $60\%$ with respect to the single-phase case. For the neutrally-buoyant fibers, instead, the macroscopic effect on the flow is always very limited. This evidence clearly suggests that the parameter governing the turbulence modulation is the mass fraction $\mathcal{M} = m_\mathrm{s}/(m_\mathrm{s}+m_\mathrm{f})$ (where $m_\mathrm{s}$ and $m_\mathrm{f}$ are the solid and fluid mass, respectively). This is reported in the inset of Fig.~\ref{fig:Kmaxc-vs-ReL} showing that the variation of $\mathrm{Re}_\lambda$ for all cases (neutrally-buoyant vs inertial, short vs long, rigid vs flexible, dilute vs dense) properly collapses. On the other hand, it can be noticed that the influence of the bending stiffness on the flow is minimal, except for very high concentrations where its effect is slightly amplified.

\begin{figure}[b]
    \centering
    \includegraphics[width=.45\textwidth]{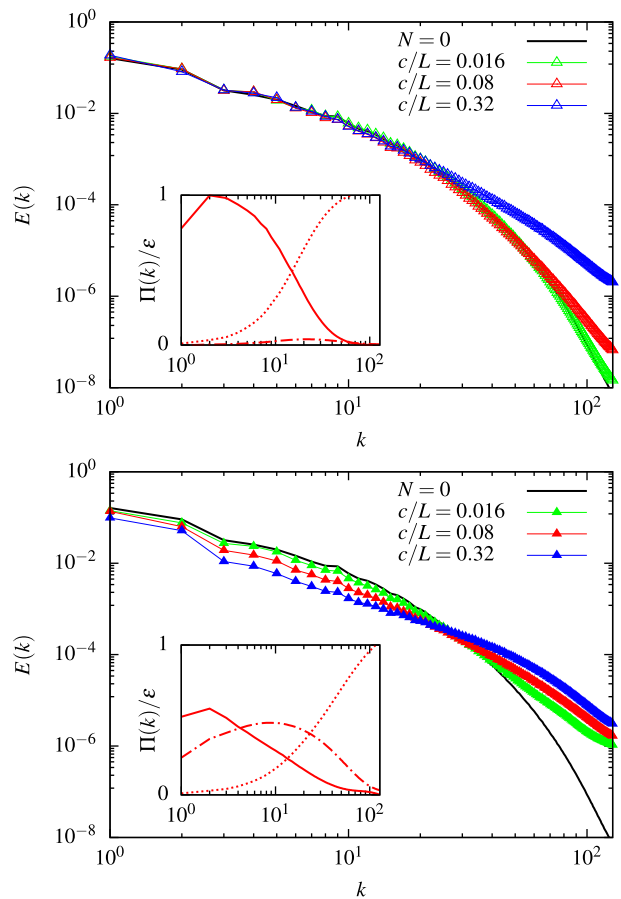}
    \caption{
    \added{Energy spectra of the modulated turbulent flow for $N=10^3$ fibers with $\gamma/\gamma_0 \approx 1$. Top: neutrally-buoyant fibers.  Bottom: inertial fibers. Colors denote different fiber length (green: short; red: intermediate; blue: long). The black solid line refers to the single-phase case, i.e. $N=0$. For the intermediate fiber length, the insets show the spectral energy fluxes associated with the nonlinear terms (solid line) and fluid-solid coupling (dot-dashed line), along with the viscous dissipation (dotted line); quantities are normalized with the energy dissipation rate. }}
    \label{fig:spectra}
\end{figure}

An improved comprehension on the backreaction effect can be obtained from the flow energy spectra reported in Fig.~\ref{fig:spectra}. 
\added{We consider the three fiber lengths of Fig.~\ref{fig:snapshots} both for the neutrally-buoyant (top) and inertial (bottom) case.}  When fibers are added, the induced backreaction can be interpreted as a combination of a large-scale ( i.e. low-wavenumber) energy subtraction and a small-scale (i.e. high-wavenumber) energy injection.  We notice that this mechanism essentially resembles the `spectral short-cut'  observed in intra-canopy flows~\cite{finnigan2000review,olivieri2020dispersed}. Overall, the decrease of total kinetic energy can be effectively described via a Darcy friction model~\cite{olivieri2020turbulence}.  A clearer picture  can be obtained by the inset of Fig.~\ref{fig:spectra}, showing (for the intermediate fiber length) the energy fluxes associated with the nonlinear term and fluid-structure coupling, along with the viscous dissipation~\cite{Note1}. For the inertial case \added{(bottom)}, the fluid-structure coupling is comparable to the nonlinear term. Conversely, in the neutrally-buoyant case \added{(top)} the former is negligible and the classical scenario for the single-phase case is recovered. For both cases, all the energy is entirely dissipated by the viscous term. Note that this is different from what happens in viscoelastic turbulence where the polymeric stresses both transfer and dissipate energy~\cite{de2005homogeneous,valente2014effect}. The fluid-structure coupling therefore purely transfers energy from larger to smaller scales, in a strict analogy with the nonlinear term; therefore, the total energy flux is given by the sum of the two contributions. Moreover, the lengthscales at which the viscous dissipation becomes effective are found to decrease in the inertial-fiber case; this is reflected in an extension of the power-law scaling region in the energy spectra (Fig.~\ref{fig:spectra}) for increasing mass fractions. 
\added{Further increasing $\mathcal{M}$ (e.g., by considering the case with the longest inertial fibers), the fluid-structure coupling becomes the dominant contribution of the flux with the nonlinear term having a negligible role (not shown). 
Despite the extended range of scaling in Fig.~\ref{fig:spectra}, the underlying balance does not imply a Kolmogorov scenario. Indeed, from the inset of Fig.~\ref{fig:spectra} it is clear that the fluid-structure interaction triggers a detectable small-scale activity and, as such, a different mechanism with respect to the classical Richardson cascade is acting to create energy-containing scales.
}
Finally, the scale-by-scale analysis confirms that no significant differences are observed when varying the bending stiffness.
 
Notwithstanding the huge variability of the analyzed configurations (both for the fluid and dispersed phase), fibers can attain only \emph{two} dynamical states. Revisiting Fig.~\ref{fig:flapFreq}, we consider the flapping frequency $f_\mathrm{flap}$ as a function of the natural frequency  $f_\mathrm{nat}$.  Following the idea of Refs.~\cite{rosti2018flexible,rosti2019flowing} in order to outline different dynamical regimes, we normalize both quantities with the hydrodynamic frequency $f_\mathrm{tur}$ associated with the turbulent eddies of size comparable to the fiber.  \added{Indeed, when using a more standard approach based, e.g., on the large-eddy turnover time, a systematic deviation between fibers of different length is found, as shown in the inset of Fig.~\ref{fig:flapFreq}.}
The straight extension of the model proposed for the dilute case, where the eddy turnover frequency \added{(at the fiber lengthscale)} is estimated using the classical Kolmogorov scaling, however, would fail due to the loss of such phenomenology when the backreaction effect is not negligible (Fig.~\ref{fig:spectra}).  Instead, we obtain the \emph{effective} hydrodynamic frequency of the modulated turbulent flow as $f_\mathrm{tur} =  \beta \, \sqrt{S_2} / c$, where $S_2 = \langle (\delta u_\parallel)^2 \rangle$ is the second-order structure function of the longitudinal velocity difference $\delta u_\parallel$ between the fiber ends and $\beta = \mathcal{O}(1)$ is a constant.  Note that similar results are found if the flow is sampled by means of Lagrangian fiber tracking instead of using the Eulerian structure function~\cite{cavaiola2019assembly,brizzolara2020fiber}.  Using either alternative approach for measuring the hydrodynamic frequency, the same outcome obtained for the dilute condition clearly emerges in Fig.~\ref{fig:flapFreq}. Considering the inertial fibers (filled symbols), two flapping states are possible: for $f_\mathrm{nat}/f_\mathrm{tur} \ll 1$, the structural response is slower than the flow forcing and consequently the fiber motion is governed by the turbulent fluctuations, i.e. lying on the horizontal plateau $f_\mathrm{flap}/f_\mathrm{tur} \approx 1$ (dotted line), whereas when $f_\mathrm{nat}/f_\mathrm{tur} \gg 1$ the restoring elastic force is faster than turbulence and controls the dynamics, leading to $f_\mathrm{flap}/f_\mathrm{tur} \approx f_\mathrm{nat}/f_\mathrm{tur}$ (dashed line).  Indeed, inertial fibers are found to lie in the so-called  underdamped regime, whilst neutrally-buoyant fibers (empty symbols) fall into the overdamped one~\cite{rosti2019flowing}. In the latter, the free-response behavior is not possible and therefore the fiber motion is always slaved to turbulence, i.e. $f_\mathrm{flap}/f_\mathrm{tur} \approx 1$.

In this Letter, we have investigated the mutual interaction between a turbulent flow and an elastic dispersed phase by means of fully-coupled direct numerical simulations.  We found that only two possible flapping states exist, one in which flexible fibers oscillate at the natural frequency and another where they synchronize to the turbulence timescales of the fiber-laden modulated flow. This universal scenario is found independently from the concentration of the suspension and the geometrical and mechanical properties of the elastic objects. The resulting flapping state is uniquely determined by the ratio between the characteristic timescales of the elastic fiber and turbulent eddies of comparable size. On the other hand, the flow modification due to the presence of the immersed objects has a complex dependency on the various suspension parameters. However, the large-scale, macroscopic behavior can be effectively described as a function of a single parameter, the mass fraction. Furthermore, a scale-by-scale analysis shows that fibers provide a non-local energy transfer, subtracting energy from the large scales and reintroducing it at smaller ones, acting in a similar way to the energy flux caused by the nonlinear terms.

\added{A final word of caution, in the case of very dense suspensions (i.e., at much higher concentrations with respect to those considered in the present study) the problem could be enriched by short-range interactions among fibers (e.g., friction and Van der Waals forces) which are not accounted in the present work}. Nevertheless, our findings are expected to hold as long as the latter are negligible compared with the hydrodynamic interaction, which is typically the case in the framework of turbulent flows.
In conclusion, an improved comprehension on the mutual interaction between fluid turbulence and elastic structures is obtained, not limited to the analysis of flexible-fiber suspensions. Indeed, the combined knowledge on the flapping states of the elastic objects and the flow alteration due to their presence, provides a complete view on the flow-structure interaction problem which can be used to effectively predict the behavior of such complex but general physical system.\\[6pt]

The interested reader is referred to the Supplemental Material containing more information on the physical model, numerical method and performed analysis.\\[6pt]

S.O. and M.E.R. acknowledge the computer time provided by the Scientific Computing section of Research Support Division at OIST. 
A.M. thanks the financial support from the Compagnia di San Paolo, project MINIERA n. I34I20000380007.\\[6pt]

The data that support the findings of this study are available from the corresponding author upon reasonable request.

%

\end{document}